\documentclass[aps, prb, twocolumn, amssymb, amsmath, showpacs, superscriptaddress]{revtex4}
\usepackage{graphicx}
\usepackage{epstopdf}
\usepackage{dcolumn}
\usepackage{bm}
\usepackage{times}
\usepackage{booktabs}
\usepackage{setspace}
\usepackage[colorlinks,
            linkcolor=red,
            anchorcolor=green,
            citecolor=blue
            ]{hyperref}

\def\eqa{\begin{eqnarray}}
\def\eea{\end{eqnarray}}
\newcommand{\eq}{\begin{equation}}
\newcommand{\ee}{\end{equation}}

\begin{document}

\title{ {\textit{d+id'}} Chiral Superconductivity in Bilayer Silicene}

\author{Feng Liu$^\star$}
\affiliation {School of Physics, Beijing Institute of Technology,
Beijing 100081, China} \affiliation {State Key Laboratory of
Nonlinear Mechanics, Institute of Mechanics, Chinese Academy of
Sciences, Beijing 100190, China}

\author{Cheng-Cheng Liu$^\star$}
\affiliation {School of Physics, Beijing Institute of Technology,
Beijing 100081, China} \affiliation {Beijing National Laboratory
for Condensed Matter Physics and Institute of Physics, Chinese
Academy of Sciences, Beijing 100190, China}

\author{Kehui Wu} \affiliation {Beijing National Laboratory for Condensed Matter Physics and Institute of Physics, Chinese Academy of Sciences, Beijing 100190, China}

\author{Fan Yang}
\thanks{\texttt{yangfan\_blg@bit.edu.cn}}
\affiliation {School of Physics, Beijing Institute of Technology,
Beijing 100081, China}

\author{Yugui Yao}
\thanks{\texttt{ygyao@bit.edu.cn}\\ $^\star$These authors contributed equally to this work.}
\affiliation {School of Physics, Beijing Institute of Technology,
Beijing 100081, China}
\affiliation {Beijing National Laboratory
for Condensed Matter Physics and Institute of Physics, Chinese
Academy of Sciences, Beijing 100190, China}

\begin{abstract}
 We investigate the structure and physical properties
of the undoped bilayer silicene through first-principles calculations
and find the system is intrinsically metallic
with sizable pocket Fermi surfaces. When realistic
electron-electron interaction turns on, the system is identified
as a chiral $d+id'$ topological superconductor mediated by the
strong spin fluctuation on the border of the antiferromagnetic spin
density wave order. Moreover, the tunable Fermi pocket area via
strain, makes it possible to adjust the spin density wave critical interaction
strength near the real one and enables a high superconducting critical
temperature.
\end{abstract}


\maketitle
 {\bf Introduction}: The chiral superconductivity (SC) is a special kind of topological
 SC characterized by time reversal  symmetry breaking~\cite{sigrist}.
 In the past few years, a surge of theoretical proposals has been raised on the
 experimental realization of this kind of unconventional SC,
 including such examples as the triplet $p_{x}\pm ip_y$ $(p+ip')$
 pairing~\cite{fu_liang,sun_kai} and the singlet $d_{x^{2}-y^{2}}\pm id_{xy}$ $(d+id')$
 pairing~\cite{laughlin,sachdev,horovitz,doniach,Kiesel,levitov}.
 While the former has probably been realized in the $Sr_{2}RuO_{4}$
 system~\cite{mackenzie}, the recent proposals\cite{doniach,Kiesel,levitov} on the realization
 of the latter in the doped graphene system have aroused
 great research interests. As a result of its nontrivial topological property, the $d+id'$ pairing state will bring a series
 of interesting experimental consequences such as quantized boundary
 current~\cite{laughlin}, spontaneous magnetization~\cite{laughlin,horovitz},
 and quantized spin and thermal Hall conductance~\cite{horovitz}.
  More interestingly, when realistic Rashba spin-orbital coupling is added to the system,
 a Majorana fermion would appear at the edge when tuning a Zeeman field\cite{majorana}.
  While the experimental realization of this intriguing pairing
 state in the system may possibly suffer from such difficulty as disorders induced by doping,
 here we predict the realization of it in another system, i.e., the
 undoped bilayer silicene (BLS) , which can avoid such difficulty.

Silicene, considered as the silicon-based counterpart of
graphene, has attracted much attention both theoretically and
experimentally~\cite{guzman-verri_electronic_2007,liu_quantum_2011,
Cahangirov_silicon_2009, ezawa_valley-polarized_2012, ezawa_quasi-topological_2012, lalmi_epitaxial_2010, feng_evidence_2012, vogt_silicene:_2012, fleurence_experimental_2012, kara_review_2012}.
 On the one hand, similar honeycomb lattice structures of the two
systems let them share most of their marvelous physical
properties, especially the gapless Dirac fermions at the
Brillouin-zone corner. On the other hand, due to the noncoplanar
low-buckled (LB) geometry in silicene, the effective first-order
spin-orbital coupling results in the quantum spin Hall effect, which
can be observed in an experimentally accessible temperature
regime~\cite{liu_quantum_2011}. Moreover, when the
exchange field and external perpendicular electric field are
added, the quantum anomalous Hall and valley polarized quantum Hall
effect can be induced~\cite{ezawa_valley-polarized_2012}.
Just like bilayer graphene (BLG), silicene can also take the
form of its bilayer version, which has recently been synthesized
\cite{feng_evidence_2012}. However, due to the LB
structure of each silicene layer, there are actually more stacking
ways between the two layers in the BLS than in the BLG. Therefore,
it is important to study the stacking structure between the
silicene bilayer and the corresponding exotic physical properties
of the BLS system.

In this Letter, we first identify the optimized crystal structure
and the corresponding electronic band structure of the BLS through
first-principles (FP) calculations. As a result, we find that the band structure of
the undoped system is intrinsically metallic with sizable Fermi
pockets, whose area is tunable via strain, which opens the door to
the formation of a superconducting state. Our further random-phase-approximation (RPA)
based study of the system reveals that the ground state of
the system is a chiral $d+id'$ pairing state, when the realistic Hubbard
interaction turns on. This superconducting pairing is mediated by
antiferromagnetic spin fluctuation on the border of the collinear spin
density wave (SDW) order identified. Furthermore, when the SDW critical interaction
strength is tuned near that of the real one via strain, the
superconducting critical temperature can be high. The exotic
chiral $d+id'$ SC in the BLS can thus be manipulated via strain,
which opens prospects for both studying the unconventional
topological SC in the new playground and for applications in
silicon-based electronics.

\begin{figure}
\includegraphics[width=3.4in]{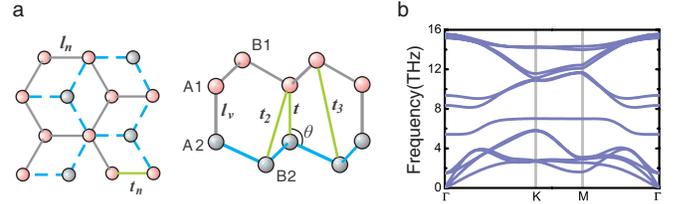}
\caption{(color online). (a) Optimized geometry of the BLS. (b) The
corresponding phonon spectrum. In (a), both the top
view (left) and side view (right) are shown. The vertical bond
length $l_{v}$, the intralayer nearest neighbor bond length $l_n$,
and the angle $\theta$ between them are marked, together with the
hopping integrals between each two of the four atoms $A_{i},
B_{i}$ $(i=1,2)$ within a unit cell.}\label{fig:crystal}
\end{figure}


{\bf Crystal and band structure}: The crystal and electronic band structures
of the BLS reported below are obtained through our FP calculations
based on the density functional theory (DFT). The electronic band
structure of the system is obtained self-consistently by using the
projector augmented wave pseudopotential method implemented
in the VASP package~\cite{kresse_efficient_1996}. The
exchange-correlation potential is treated by
the Perdew-Burke-Ernzerhof potential~\cite{perdew_generalized_1996}.

As a consequence of the LB structure of each silicene layer, there are actually four stacking
ways (see the Supplemental Material, Note
\uppercase\expandafter{\romannumeral1}) between the upper and
lower layers in the system. Our FP calculations reveal that two
of them are stable, among which the energetically favored one
(named the $AB$-$bt$ structure) is shown in Fig.\ref{fig:crystal}(a),
and the corresponding phonon
spectrum~\cite{togo_first-principles_2008} is shown in
Fig.\ref{fig:crystal}(b).

From Fig.\ref{fig:crystal}(a), the bottom ($A_1$ sublattice) of the upper silicene
layer couples with the top ($A_2$ sublattice) of the lower layer
vertically with a bond length $l_{v}=2.53$ \AA, while the two
sublattices ($A$ and $B$) within a layer couple with a bond length
$l_{n}=2.32$ \AA. Approximately equal bond lengths, together with
the bond angle $\theta=106.60^{o}$ between the two bonds describe
an orbital hybridization more like the $sp^{3}$ type (with bond
angle $\theta=109.47^{o}$) than the planar $sp^{2}$ type. From
Fig.\ref{fig:crystal}(b), the phonon frequencies obtained are real at
all momenta, which suggests a stable structure. The energy of this
configuration is -19.65 eV per unit cell, lower than that of the
configuration studied in the
literature~\cite{morishita_first-principles_2010}, which is
-19.51 eV per unit cell. It is noting here that the
symmetry group of the system is $D_{3d}$.

\begin{figure}
\includegraphics[width=3.4in]{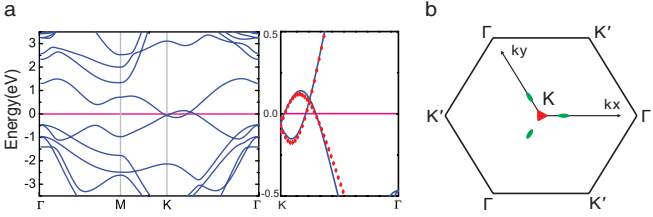}
\caption{(color online). (a) The band
structure of BLS corresponding to the optimized lattice structure shown in Fig.\ref{fig:crystal}(a).
(b) The FS patches around the $K$ point. In (a),
the zooming in low energy band (right) is also shown, where the tight-binding (TB)
model (red scatters) is compared with the FP results (blue solid line).
In (b), the central pocket (red) is electronlike and the
outer three identical pockets (green) are holelike,
with the total areas of the two kinds of pockets equal.}\label{fig:electronic}
\end{figure}
The band structure of the BLS with the $AB$-$bt$ stacking way is shown in Fig.\ref{fig:electronic}(a) (left), together with
its low energy zooming in (right). The most obvious feature of
this band structure is the 300 meV overlap between the valence band
and the conduction band, much larger than the 1.6 meV in the BLG
and the 40 meV in the graphite \cite{mccann_landau-level_2006,partoens_graphene_2006}. Another important feature is the
band crossings present not only at the $K$ points, but
also on the $K$-$\Gamma$ axes with an energy difference between
them. Such band crossings result in a three-folded
symmetric pocket Fermi surface (FS) structure surrounding each $K$
point, as shown in Fig.\ref{fig:electronic}(b), where the central
electron pocket is accompanied by three identical outer hole
pockets with equal total area. Here, only the FS patches around
one $K$ point are present. The other patches can be obtained by
six-folded rotations around the $\Gamma$ point, as required by the
$D_{3d}$ symmetry and the time reversal invariant of the system.


To proceed, we construct the following effective
four-band TB model in the basis
$\{|B_1\rangle,|B_2\rangle,|A_1\rangle,|A_2\rangle\}$, which well
captures all the low energy features of the above band structure
near the FS [compared with FP results in Fig.\ref{fig:electronic}(a)],
\begin{eqnarray}
\hspace{-5mm}
H(\mathbf{k})=\left(
\begin{array}{cccc}
\Delta        & t_{3}f(\mathbf{k})    & t_{n}f(\mathbf{k})^* &-t_{2}f(\mathbf{k})^*\\
t_{3}f(\mathbf{k})^*  & \Delta       &-t_{2}f(\mathbf{k})   & t_{n}f(\mathbf{k})\\
t_{n}f(\mathbf{k})    &-t_{2}f(\mathbf{k})^*  & 0           & t\\
-t_{2}f(\mathbf{k})    &t_{n}f(\mathbf{k})^*  & t           & 0\\
\end{array}
\right).\label{TB}
\end{eqnarray}
Here, $A_{i},B_{i}$ $(i=1,2)$ represent the basis mainly composed of
the $3p_z$ orbitals localized around each of the four silicon
atoms  within a unit cell. The hopping integrals $t_{n}$, $t$,
$t_2$, and $t_3$ between each two orbitals are marked in
Fig.\ref{fig:crystal}(a). The phase factor $f(\mathbf{k})
$ is $\sum_{\alpha}e^{i{\mathbf{k}} \cdot {{\mathbf{R}}_{\alpha}}}$,
with $\mathbf{R}_{\alpha}$ ($\alpha=1,2,3$) to be the
nearest-neighbor vector. Finally, notice the small effective on-site
energy difference $\Delta$ between the $A$ and $B$ atoms. The
fitted parameters of the system in comparison with those of the
BLG are listed in Ref.\cite{TB_compare}, from which the
most obvious feature of the BLS lies in the dominating role of the
vertical interlayer hopping $t$. The resulting strong
bonding-antibonding energy split between the $A_1$ and $A_2$ orbitals
pushes them far away from the Fermi level, leaving the $B_1$ and $B_2$
orbitals to form a low energy subspace which takes responsibility for the
main physics of the system.

\begin{figure}
\includegraphics[width=3.4in]{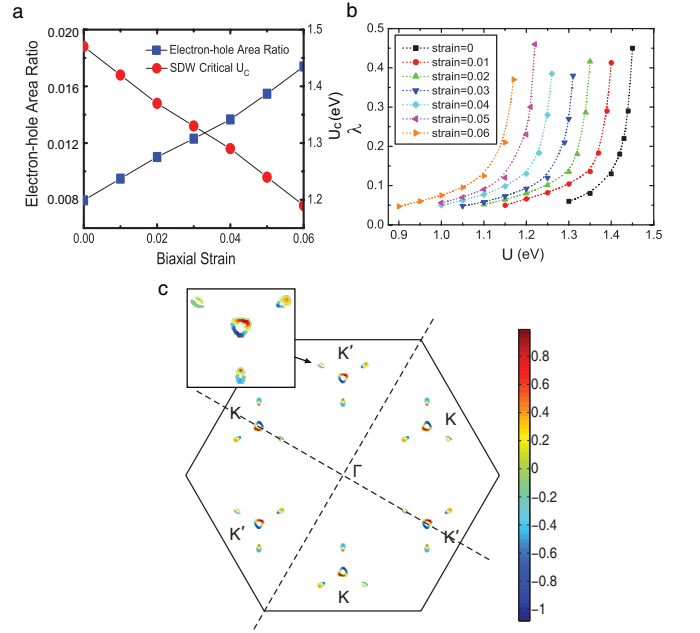}
\caption{(color online). (a) The biaxial strain dependence of Fermi pocket area
ratio viz. the ratio of the total area of the electron and hole
pockets against the total area of the Brillouin zone and the SDW
critical value $U_c$ of the BLS. (b) The interaction
strength $U$ dependence of the largest eigenvalue $\lambda$ of the
linearized gap function (\ref{gap_function}), which is related to
$T_c$ through $T_{c}=1.13 \hbar\omega_{D}e^{-1/\lambda}$. Results
for different strain values are compared. (c) $\mathbf{k}$ dependence
of the gap function of the $d_{x^{2}-y^{2}}$ symmetry for $U$=1 eV, which is
antisymmetric about the axes $x=\pm y$ shown in the
reciprocal space.  Inset: zooming in the vicinity of
$K'$.}\label{fig:sc}
\end{figure}

It is important to point out here that the low energy band
structure of the system is considerably sensitive to biaxial
strains. As shown in
Fig.\ref{fig:sc}(a), for small strains which keep the symmetry and
FS topology of the system, the total area of the electron or hole
pockets feels a considerable variation. This tunable property of
the band structure turns out to be very important for our
following discussions.

\begin{figure}
\includegraphics[width=3.4in]{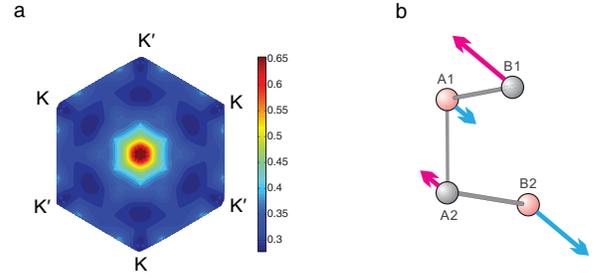}
\caption{(color online). (a) $\mathbf{k}$ dependence of the free static
susceptibility. (b) The SDW ordered spin pattern within a unit cell.}\label{fig:SDW}
\end{figure}

{\bf Model and free susceptibility}: Let us consider the following
four-band Hubbard-model of the system:
\begin{equation}
H=\sum_{\mathbf{k}\sigma,\alpha\beta}c_{\mathbf{k}\alpha\sigma}^
{\dagger}H_{\alpha\beta}(\mathbf{k})c_{\mathbf{k}\beta\sigma}+
U\sum_{i,\alpha=1,4}n_{i\alpha\uparrow}n_{i\alpha\downarrow},\label{model}
\end{equation}
where $H(\mathbf{k})$ is defined by Eq.(\ref{TB}), $i$ and $\alpha$
($\beta$) denote the unit cell and orbital indices, respectively.
Noticing that the electron-electron interaction has been included
in the mean-field level in our DFT calculation, the explicit inclusion
of the Hubbard repulsion here would lead to a slight renormalization
of the TB parameters. However, such a slight parameter renormalization
would not qualitatively change the main physics here.

The free susceptibility $\chi_{l_{3},l_{4}}^{(0)l_{1},l_{2}}\left(\mathbf{q},i\omega_{n}\right)$ (for $U=0$) of the model is given in
Supplemental Material, Note \uppercase\expandafter{\romannumeral2},
and the $\mathbf{k}$-dependent static susceptibility of the system
defined by the largest eigenvalue of the susceptibility matrix
$\chi^{(0)}_{l,m}\left(\mathbf{q}\right)\equiv
 \chi^{(0)l,l}_{m,m}\left(\mathbf{q},i\nu=0\right)$ is shown in
 Fig.\ref{fig:SDW}(a), which displays a distribution centering around
 the $\Gamma$ point. Note that the susceptibility peak at the $\Gamma$ point only suggests the same repeating pattern from one unit cell to another and within one single unit cell there can be antiferromagnetic structure, as introduced below.

{\bf SDW and SC}: When the Hubbard interaction turns
on, the standard multiorbital RPA \cite{rpa1,rpa2,scalapino} (see
also Supplemental Material \uppercase\expandafter{\romannumeral2}) approach is adopted in our study. The spin [$\chi^{(s)}$]
or charge [$\chi^{(c)}$] susceptibilities in the RPA level are defined
in Supplemental Material, Note \uppercase\expandafter{\romannumeral2},
and it is clear that the repulsive Hubbard interaction suppresses $\chi^{(c)}$ (hence, charge density wave instability) and
enhances $\chi^{(s)}$ (hence, SDW instability). When the interaction strength $U$ is
enhanced to a critical value $U_{c}$, the spin susceptibility of
the model diverges, which implies the instability toward
long-range SDW order. The ordered spin structure of this bilayer
system determined by the eigenvector of the spin susceptibility
matrix $\chi^{(s)}_{l,m}\left(\mathbf{q}\right)\equiv
 \chi^{(s)l,l}_{m,m}\left(\mathbf{q},i\nu=0\right)$ corresponding to
its largest eigenvalue is shown in Fig.\ref{fig:SDW}(b), from which
one finds an antiferromagnetic state with antiparallelly aligned
spin patterns within a unit cell. The ordered moments are mainly
distributed on the $B_{i}$ $(i=1,2)$ atoms which take
responsibility for the low energy physics of the system. It is
noting here that with the enhancement of the strain and hence the
Fermi pocket area, the SDW critical value $U_c$ feels an obvious
variation from the 1.48 eV at zero strain to the 1.18 eV at the
strain of 0.06. Such a range is probably realizable for the
Hubbard $U$ of the $3p_z$ orbitals of the silicon atoms.

When the Hubbard $U$ is near but lower than $U_c$, the
antiferromagnetic spin fluctuation is strong in the system.
Through exchanging such antiferromagnetic spin
fluctuations between each Cooper pair, unconventional chiral
$d+id'$ SC emerges in the BLS system.

The effective interaction obtained in the RPA level generated through exchanging spin susceptibility is $V_{eff}=
 \sum_{\alpha\beta,\mathbf{kk'}}V^{\alpha\beta}(\mathbf{k,k'})c_{\alpha}^{\dagger}(\mathbf{k})
 c_{\alpha}^{\dagger}(-\mathbf{k})c_{\beta}(-\mathbf{k}')c_{\beta}(\mathbf{k}')$, (see Supplemental Material, Note \uppercase\expandafter{\romannumeral2}), from which one obtains the following linearized gap
 equation \cite{scalapino} near $T_c$:
 \begin{equation}
 -\frac{1}{(2\pi)^2}\sum_{\beta}\oint_{FS}
dk'_{\Vert}\frac{V^{\alpha\beta}(\mathbf{k,k'})}{v^{\beta}_{F}(\mathbf{k'})}\Delta_{\beta}(\mathbf{k'})=\lambda
 \Delta_{\alpha}(\mathbf{k}).\label{gap_function}
\end{equation}
Here, the integration is along the $\beta$ th FS patch. The $v^{\beta}_{F}(\mathbf{k'})$ is the Fermi
velocity, and $k'_{\Vert}$ represents the component along the FS. Diagonalizing this eigenvalue problem, one obtains the largest eigenvalue $\lambda$, which is related to the $T_c$ of the system through $T_{c}=1.13
\hbar\omega_{D}e^{-1/\lambda}$, and the corresponding eigenvector $\Delta_{\alpha}(\mathbf{k})$ which determines the leading pairing symmetry of the system. Here, $\hbar\omega_D$ is a typical
energy scale for the spin fluctuation, approximated as
the low energy bandwidth, i.e., $\hbar\omega_D\approx 300$ meV.

Our RPA calculations on the BLS identify exactly degenerate
$d_{xy}$ and $d_{x^{2}-y^{2}}$ doublets as the leading pairing
symmetries of the system for $U<U_c$ at all strain values, which is
robust against small doping (see Supplemental Material, Note
\uppercase\expandafter{\romannumeral3}), as can be induced by the supporting substrate\cite{Profeta}. Both symmetries are
singlet with nodal gap functions. While the $d_{x^{2}-y^{2}}$
shown in Fig.\ref{fig:sc}(c) is antisymmetric about the axis $x=\pm
y$ in the reciprocal space, the $d_{xy}$ shown
in Supplemental Material, Note \uppercase\expandafter{\romannumeral3} is symmetric about them. The two gap functions
form a 2D $E_g$ representation of the $D_{3d}$ point group of the
system. For both symmetries, two gap nodes are present on each
Fermi pocket.

Since the two $d$-wave pairing symmetries are degenerate here in
the quadratic level of the Ginsberg-Landau free energy, they would generally
be superposed \cite{levitov} as $\Delta\left(\mathbf{k}\right)=
\Delta_{1}\Delta_{d_{x^{2}-y^{2}}}\left(\mathbf{k}\right)+
\Delta_{2}\Delta_{d_{xy}}\left(\mathbf{k}\right)$ to further
lower the energy up to higher levels. Our energy optimization on the
effective Hamiltonian $H_{eff}=H_{band}+V_{eff}$ yields
$\Delta_2=i\Delta_1$, which just leads to the long-sought nodeless
chiral $d+id'$ SC. This superposition manner between the two
$d$-wave symmetries satisfies the requirement that the gap nodes
should avoid the FS to lower the energy. With intrinsically
complex gap function, this pairing breaks time reversal symmetry
and will bring a lot of exotic properties. It is a singlet
analogy of the extensively studied $p+ip'$ SC.

The $U$ dependence of the eigenvalue $\lambda$ of
Eq.(\ref{gap_function}) which is related to $T_c$, is shown in
Fig.\ref{fig:sc}(b) for different strains. Clearly, $T_c$ increases
with the Hubbard $U$ and rises promptly at $U/U_{c}\lesssim 1$ as a
result of the strongly enhanced antiferromagnetic spin fluctuation
near the critical point. Since $U_c$ is tunable via strain, as
shown in Fig.\ref{fig:sc}(a), the ratio $U/U_{c}$ varies within a
range which provides basis for the realization of the relation
$U/U_{c}\lesssim 1$ which is crucial for the high $T_ c$ of the
system. For example, for $\lambda\approx0.3$ attainable by
different strains shown in Fig.\ref{fig:sc}(b), the $T_c$ obtained
can be as high as $80$ K or more, although it is usually overestimated
in the RPA level. For real material, whether high $T_c$ can be
acquired is determined by how near $U/U_c$ can be tuned to
1.

Our RPA calculations for the system also identify a possible
nodeless $f$-wave pairing to be the leading symmetry in the triplet
channel, consistent with Ref.\cite{Roy_arXiv_2012}. This pairing also breaks time reversal symmetry and the
gap function changes sign with every $60^{o}$ rotation, which
belongs to the $A_{1u}$ representation of $D_{3d}$ (see Supplemental Material,
Note \uppercase\expandafter{\romannumeral3}).
However, its $T_{c}$ is much lower than that of the $d+id'$
pairing. Note that there are also discussions on the competition
among various superconducting symmetries in graphene\cite{Pellegrino}.

The $d+id'$ pairing symmetry obtained here is
reliable, as it is based on the weak coupling RPA approach. As for
the superconducting critical temperature $T_c$, the RPA approach
generally overestimates $T_c$ near the critical point. Thus, the
$T_c$ in real material might be lower than that estimated here.
How to calculate the $T_c$ accurately remains an open
question. What is more, the coexistence between the SDW order and
SC is also possible in the system, which is beyond the present
framework. Furthermore, other types of many-body ordered states discussed in the BLG
\cite{Vafek}are also possible here. Further studies are needed for such purposes.

{\bf Conclusion}: We have performed a FP calculation on the BLS.
Through energy optimization, we identified a $D_{3d}$ symmetric
stacking structure for the system. The band structure
corresponding to this crystal structure was intrinsically metallic,
with Fermi pockets around each $K$ point whose areas were tunable via
strain. Further RPA-based studies predicted a chiral
$d+id'$ superconducting ground state of the system for realistic
electron-electron interactions.  The superconducting critical
temperature of this spin-fluctuation mediated SC was
well tunable via strain, which could be high when the SDW
critical interaction strength was tuned near that of the real one.
The realization of the chiral $d+id'$ SC in the BLS predicted here
will not only provide a new playground for the study of the
topological SC, but also bring a new epoch to the familiar
Si industry.

\textbf{Acknowledgement}: We are grateful to Dung-Hai Lee, Jun-Ren Shi,
Fa Wang, and Hong Yao for stimulating discussions. The
work is supported by the MOST Project of China (Grants
No. 2014CB920903, No. 2011CBA00100) and the NSF of
China (Grants No. 10974231, No. 11174337, No. 11274041,
and No. 11225418). F. Y. is supported by the NCET program
under Grant No. NCET-12-0038. F. L. and C.-C. L. contributed equally to this work. \\

\end{document}